\pdfoutput=1
\documentclass[a4paper]{llncs}

\usepackage{graphicx}
\usepackage{amsmath}
\usepackage{amsfonts}
\usepackage[T1]{fontenc}
\usepackage{varioref}
\usepackage{xspace}
\usepackage{paralist}
\usepackage{fancybox}
\usepackage{xcolor}
\usepackage{calc}
\usepackage{verbatim}

\usepackage{relsize}
\usepackage{booktabs}
\usepackage[scaled]{helvet}

\usepackage{pdfsetup}
\usepackage{times}

\usepackage{listingsVDM}

\usepackage{overturelanguagedef}

\definecolor{mGreen}{rgb}{0,0.6,0}
\definecolor{mGray}{rgb}{0.5,0.5,0.5}
\definecolor{mPurple}{rgb}{0.58,0,0.82}
\definecolor{backgroundColour}{rgb}{0.95,0.95,0.92}

\definecolor{maroon}{rgb}{0.5,0,0}
\definecolor{darkgreen}{rgb}{0,0.5,0}
\definecolor{ao}{rgb}{0.0, 0.5, 0.0}
\definecolor{mycolor1}{rgb}{0.0, 0.53, 0.74}%
\definecolor{mycolor2}{rgb}{0.21783,0.72504,0.61926}%
\definecolor{mycolor4}{rgb}{0.93, 0.53, 0.18}%
\definecolor{plum}{rgb}{0.56, 0.27, 0.52}
\definecolor{pinegreen}{rgb}{0.0, 0.47, 0.44}
\definecolor{pthaloblue}{rgb}{0.0, 0.06, 0.54}
\definecolor{saffron}{rgb}{0.96, 0.77, 0.19}

\usepackage{cleveref}

\usepackage[caption=false]{subfig}
\usepackage{pgfplots}
\pgfplotsset{compat=newest}
\usetikzlibrary{pgfplots.groupplots, pgfplots.units}
\pgfplotsset{
	every axis label/.append style={font=\normalsize},
	tick label style={font=\small},
	/pgfplots/enlargelimits=false,
    legend style={legend pos=north east, font=\small},
    legend cell align=left,
    xlabel near ticks,
    ylabel near ticks,
	axis on top,
    highlight/.code args={#1:#2}{
        \fill [every highlight] ({axis cs:#1,0}|-{rel axis cs:0,0}) rectangle ({axis cs:#2,0}|-{rel axis cs:0,1});
    },
    /tikz/every highlight/.style={
        on layer=\pgfkeysvalueof{/pgfplots/highlight layer},
        red!10
    },
    /tikz/highlight style/.style={
        /tikz/every highlight/.append style=#1
    },
    highlight layer/.initial=axis background
}%

\lstdefinelanguage{XML}
{
  basicstyle=\ttfamily\scriptsize,
  morestring=[b]",
  moredelim=[s][\bfseries\color{maroon}]{<}{\ },
  moredelim=[s][\bfseries\color{maroon}]{</}{>},
  moredelim=[l][\bfseries\color{maroon}]{/>},
  moredelim=[l][\bfseries\color{maroon}]{>},
  morecomment=[s]{<?}{?>},
  morecomment=[s]{<!--}{-->},
  commentstyle=\color{darkgreen},
  stringstyle=\color{blue},
  identifierstyle=\color{red}
}

\newcommand{\ie}{i.e., }

\usepackage{tikz}
\usepackage{balance}
\usetikzlibrary{shapes,fit,arrows,calc,arrows,arrows.meta,positioning}

\begin{document}

\title{QuickCheck for VDM}
\titlerunning{QuickCheck for VDM}

\author{Nick Battle\inst{1} \orcidID{0009-0001-1523-4964} \and
        Markus Solecki Ellyton\inst{2} \orcidID{0009-0000-4606-5019}
        }
\institute{School of Computing, Newcastle University, UK \email{nick.battle@ncl.ac.uk}
\and
DIGIT, Department of Electrical and Computer Engineering, Aarhus University, Denmark, \email{202302976@post.au.dk}
}

\maketitle

\begin{abstract}
We describe recent work on a lightweight verification tool for VDM specifications, called \emph{QuickCheck}. The objective of the tool is to quickly categorise proof obligations: identifying those that fail with counterexamples, those that are probably provable and those that require deeper analysis. The paper discusses the design of the tool and its use of pluggable strategies for adding extra checking. We present the results of the tool being used to check a large set of VDM specifications, and suggest future directions.
\end{abstract}

\section{Introduction}
\label{sec:intro}
Tools for working with VDM specifications have been able to produce \emph{proof obligations}\footnote{VDMTools~\cite{Larsen01} calls them "integrity properties"} for many years. A proof obligation is a small VDM-SL boolean expression, which should always be true - it should be a tautology. If this can be proved, then some aspect of the specification has been verified, for example showing that a calculation never divides by zero.

So formal verification of VDM specifications is centered on the proof or discharge of proof obligations. However, until recently, tool support for the automatic discharge of proof obligations was missing.

This situation improved with the introduction of the Isabelle translation plugin for VDMJ~\cite{Battle09,FJVW13TR}. With this plugin, both the original specification and its proof obligations are translated into Isabelle/HOL, such that the Isabelle prover can then attempt to automatically discharge them.

Although this process is very powerful, it is also very sophisticated, requiring the user to have a thorough understanding of both the VDM and Isabelle languages. If issues are encountered with the translation or the discharge of obligations in Isabelle, these have to be corrected in the VDM source. The workflow is also not trivial, requiring the user to switch between tool environments, for example.

An ideal proof system would be tightly integrated with the development tool, so that proof obligations are generated and discharged almost without the user being aware of them, unless issues are discovered. Of course this is difficult to achieve, but the objective of the \emph{QuickCheck} plugin is to make some progress in this direction.

\emph{QuickCheck} attempts to assist the user by \emph{very quickly} generating proof obligations and labelling them with one of three main categories\footnote{See \url{https://github.com/nickbattle/vdmj/wiki/Using-QuickCheck}}. Firstly, if a counterexample can be found quickly (an assignment of variables that makes the obligation expression false), then this is displayed as a warning in the editor and it allows the counterexample to be debugged. Secondly, if we can quickly determine that an obligation is very likely to be a tautology, then we can indicate that it is \emph{probably provable} via a theorem prover. Lastly, if neither of these is the case, we can indicate that we are not sure whether the obligation is provable and therefore that a theorem prover should be tried, with care.

Note that the only \emph{certain} output of the \emph{QuickCheck} plugin is when a counterexample can be found. The other two categories are not certain without theorem prover confirmation, but the indication of "probably provable" is still helpful - for example, these obligations could be discharged at the start easily, leaving the user to focus effort on the harder cases.

Section~\ref{sec:qc} looks at the design of \emph{QuickCheck} and the standard strategies included with the tool. Section~\ref{sec:integration} looks at how the tool is integrated with the VDMJ and VDM-VSCode environments. Section~\ref{sec:related} looks at related work, and finally Section~\ref{sec:future} considers future directions.

\section{Analysing Proof Obligations with QuickCheck}
\label{sec:qc}
\emph{QuickCheck} tries to categorise proof obligations by directly evaluating obligation expressions. First we give some example proof obligations and show how \emph{QuickCheck} processes them. Then we describe how this is achieved with pluggable strategies. The objective is to categorise POs as either \emph{FAILED} without a counterexample, probably \emph{PROVABLE} or  \emph{MAYBE}, if we cannot decide.

\subsection{Proof Obligations}
\label{sec:qc:pos}

As described above, a proof obligation is a small VDM-SL boolean expression which must always be true if some aspect of the specification is to always have defined behaviour. This analysis complements ad-hoc execution testing and combinatorial testing.

For example, consider a simple function that is intended to return the item at a particular index in a sequence:

\begin{verbatim}
functions
    itemAt: seq of nat * nat -> nat
    itemAt(list, index) == list(index);

> pog
Generated 1 proof obligation:

Proof Obligation 1: (Unproved)
itemAt: sequence apply obligation in test.vdmsl at line 3:28
(forall list:seq of nat, index:nat &
  index in set inds list)
\end{verbatim}

\noindent
This specification generates a single "sequence apply" proof obligation, which effectively says, \textit{for all possible argument combinations, the index is within the set of indexes of the sequence passed}. This obligation is required since, if the condition is not true, the value of \textit{list(index)} is not defined (the index is out of range).

A casual consideration of the obligation quickly reveals that it is not always true (\ie it is not a tautology), since it is easy to pass an invalid index to the list, such as an index of 4 with a sequence of \texttt{[1,2,3]}. This means that the specification should be tightened somehow to either deal with these illegal cases or prohibit them. The simplest solution is to add a precondition which states that the arguments are legal, as follows:

\begin{verbatim}
functions
    itemAt: seq of nat * nat -> nat
    itemAt(list, index) == list(index)
    pre index in set inds list;

> pog
Generated 1 proof obligation:

Proof Obligation 1: (Unproved)
itemAt: sequence apply obligation in test.vdmsl at line 3:28
(forall list:seq of nat, index:nat &
    pre_itemAt(list, index) => index in set inds list)
\end{verbatim}

\noindent
Note that now, there is a guard condition in the obligation, which calls the precondition of the function. The call to \emph{pre\_itemAt} is a total function, guaranteed to return true or false, depending on whether or not the precondition is met. This effectively eliminates the illegal argument combinations from the obligation and it is now a tautology (by casual observation - this still needs to be proved).

In more realistic examples, it is not easy to determine whether an obligation is true just by casual observation. This is where tools like \emph{QuickCheck} are helpful.

\subsection{Running QuickCheck}
If we run \emph{QuickCheck}\footnote{We show the VDMJ command line version here for ease of inclusion in the text, but see Section~\ref{sec:vscode} for the graphical version.} on the first version of the specification above (without the precondition), we get the following result:

\begin{verbatim}
> qc
PO #1, FAILED in 0.013s: Counterexample: index = 0, list = []
----
itemAt: sequence apply obligation in test.vdmsl at line 3:28
(forall list:seq of nat, index:nat &
  index in set inds list)

> qr 1
=> print itemAt([], 0)
Error 4064: Value 0 is not a nat1 in test.vdmsl at line 3:28
3:      itemAt(list, index) == list(index)

\end{verbatim}

\noindent
The \emph{FAILED} status indicates that the "qc" execution fails, very quickly. It also gives a counterexample which lists values of the variable bindings that produces a false result.

After the "qc" execution, we can run "qr" on obligation \#1, which attempts to evaluate the function enclosing the proof obligation, using the counterexample bindings to deduce the argument values to pass (though this is not always possible). This produces a runtime error, which is the "problem" that the proof obligation was checking for.

On closer examination, this tells us that the evaluation failed because the \texttt{nat} argument of zero is not valid for the \texttt{nat1} index of a sequence - \ie the type of the \texttt{index} argument is incorrect as well as there being a missing precondition.

If we change to use the version of the function with the precondition, we get the following:

\begin{verbatim}
> qc
PO #1, MAYBE in 0.023s
> 
\end{verbatim}

\noindent
The \emph{MAYBE} status means that the validity of the new obligation could not be determined. \emph{QuickCheck} could not find a counterexample, so the obligation may be valid, but we are not certain. Note that the execution time is still relatively quick.

If we change the specification to put an explicit test in the body of the function, rather than using a precondition, we get a slightly different obligation and \emph{QuickCheck} result:

\begin{verbatim}
functions
    itemAt: seq of nat * nat -> nat
    itemAt(list, index) ==
        if index in set inds list
        then list(index)
        else 0;

> pog
Generated 1 proof obligation:

Proof Obligation 1: (Unproved)
itemAt: sequence apply obligation in test.vdmsl at line 5:14
(forall list:seq of nat, index:nat &
  ((index in set (inds list)) => index in set inds list))

> qc
PO #1, PROVABLE by trivial index in set (inds list) in 0.001s
>
\end{verbatim}

\noindent
Note that the proof obligation now effectively says, \textit{if the index is valid then the index is valid}. This is a very common pattern in proof obligations, and there is a particular strategy (below) to deal with these common cases, called the \textit{trivial} strategy. When such a case is identified, the result is \emph{PROVABLE} - meaning that a theorem prover ought to be able to prove this, though strictly speaking we are not certain.

\subsection{QuickCheck Plugin Design}
\label{sec:design}

As stated above, \emph{QuickCheck} determines the validity or otherwise of a proof obligation by directly evaluating the obligation expression.

Most proof obligations include nested \texttt{forall} or \texttt{exists} expressions that quantify over type binds\footnote{For example, \texttt{forall a:nat \&...} means for every "a" of type "nat"...}. Type binds cannot be evaluated by the VDMJ interpreter if the types concerned are infinite, but it is very common to have obligations that reason about infinite types. So \emph{QuickCheck} has to solve three problems: firstly, how to enable the interpreter to evaluate type binds with a finite subset of infinite types; secondly, how to select a finite subset of values from an infinite type to test; and thirdly, how to interpret the evaluation result, to correctly indicate whether the obligation may be provable or provide a counterexample.

The first problem was relatively easy to solve. The VDMJ interpreter was instrumented in such a way that \texttt{forall} and \texttt{exists} expressions can be given sets of values for any type binds that they contain. If any evaluations return false, the bindings concerned are saved to potentially be used as counterexamples (if the overall result is false).

The third problem was also relatively easy to solve, though care has to be taken with complex nested expressions and error cases. If \emph{QuickCheck} cannot determine the status with confidence, the \emph{MAYBE} status is returned. If an evaluation takes too long, a \emph{TIMEOUT} status is given. If an obligation cannot be executed\footnote{This happens for obligations generated from within VDM operations, where there is insufficient information about the state of the system.}, an \emph{UNCHECKED} status is given.

So the most complex part of \emph{QuickCheck} is the selection of binding values that attempt to find counterexamples. There are a variety of different ways to do this, each being a compromise between trying very many values that are generated without sophistication, and trying comparatively few values, selected with more effort.

Because there are different ways to select values, the \emph{QuickCheck} design allows for pluggable "strategies" to be created and added by users, and the tool is delivered with a set of strategies built-in. Several strategies may be combined in a single run of \emph{QuickCheck}.

\smallskip
\noindent
Every strategy is passed the following information:

\begin{itemize}
    \item The proof obligation, including its expression and its location in the specification
    \item A list of type binds in the obligation to generate values for
    \item An execution Context to allow things to be evaluated during generation, like type invariants.
\end{itemize}

\noindent
And every strategy can return the following information:

\begin{itemize}
    \item A map of binding variable names to lists of possible values
    \item A \textit{hasAllValues} flag indicating whether all bindings have all possible values for their types (\ie that all types are finite and fully populated)
    \item A (dis)proved indicator
\end{itemize}

\noindent
\emph{QuickCheck} applies all of the configured strategies to each proof obligation, combining all of the generated values into one binding map. If no strategy claims to have (dis)proved the obligation, the combined bindings are used to instrument the obligation expression which is then evaluated to see whether it is true, false, or generates an error.

The overall status is derived from the result of the evaluation, whether the obligation is a \texttt{forall} or \texttt{exists} expression, whether a strategy claims to have (dis)proved the obligation and whether the \textit{hasAllValues} flag is set.

\subsection{Built-in Strategies}
\label{sec:strategies}

As mentioned above, \emph{QuickCheck} uses pluggable strategies to generate bindings to test, and the tool comes with a set of strategies built-in, as follows:

\begin{itemize}
    \item \emph{The fixed strategy.} This is the simplest strategy, which by default will generate a fixed set of values for each possible VDM type. The value generation process starts with primitive types (\texttt{nat}, \texttt{int}, \texttt{bool}, \texttt{char} and so on). These generate a fixed set of values of the type, with the numeric values including zero, if possible (zero is a value that often causes problems, hence counterexamples). For example, if 100 integers were generated, the strategy would produce \texttt{\{-50, ..., 49\}}. Then compound types (sets, sequences, records, tuples etc.) are generated using fixed combinations of the values of the constituent fields (subject to any type invariants that may apply).

    \smallskip
    An alternative option for the fixed strategy is to use a small configuration file to define explicit value sets for each binding. Lines of the configuration file are of the form \texttt{<type bind> = <set expression>}, which evaluates the set expression and assigns the values generated to the type bind, when this name/type occurs in any obligations. This is useful to generate very specific sets of values that may be relevant to the testing of a particular specification.

    \smallskip
    \item \emph{The random strategy.} This strategy generates primitive values using a pseudo-random number generator. Numeric values are strongly biased to be around zero (since counterexamples with small values are easier to understand) though progressively larger values are used as more and more are generated. For example, the first integer is in the range -10 to +10, the next is -20 to +20, and so on. As with the \emph{fixed} strategy, values of compound types are generated using combinations of the values of the component fields, but using pseudo-random selection.

    \smallskip
    By default, the pesudo-random number generator is seeded from the system clock, but a seed can be provided to give repeatable random values.

    \smallskip
    \item \emph{The trivial strategy.} This strategy takes advantage of the fact that proof obligations are often of a very predictable form, and trivially true. For example, a specification will often guard against a condition with a test (like \texttt{x<>0}) before performing an evaluation that depends on that condition (like \texttt{1/x}). The proof obligation for the division will therefore be something like \texttt{x<>0 => x<>0}. This is trivially true, so this strategy looks for these common patterns and immediately concludes that the obligation is probably provable.

    \smallskip
    \item \emph{The finite strategy.} This strategy looks for bindings where the type concerned has a finite number of values (and not too many of them). For example, \texttt{set of bool} has four possible values. In this case, instead of selecting a subset of the values of the type, the strategy can systematically generate all values of the type, and pass back the \textit{hasAllValues} flag.

    \smallskip
    This is important because if all bindings in an obligation have all values checked, and none of the combinations cause the obligation to be false, the obligation can be claimed as probably provable.

    \smallskip
    \item \emph{The search strategy.} This strategy looks for VDM boolean sub-expressions in the obligation, where a single variable is compared to a constant via an operator (for example \texttt{x > 0}). In these cases, it is easy to generate a a single bind value that would cause the expression to be false (in this example, setting \texttt{x = 0}).

    \smallskip
    Note that the strategy takes no account of the context in which the boolean sub-expression occurs, so it is very naive, but extremely fast. If it only helps to find a counterexample in rare cases, that is worth spending a microsecond or two calculating the value.
    
    \smallskip
    \item \emph{The direct strategy.} This strategy takes a different approach to the others, in that it ignores the obligation expression. Instead, it uses the kind of obligation and its location information to decide what the obligation is trying to verify. The strategy then verifies the same thing via "direct" means.

    \smallskip
    For example, a cases-exhaustive obligation tries to verify that the patterns in a \texttt{cases} expression that has no \texttt{others} clause will match every possible value passed. The proof obligation expression states this, but the form of the obligation is quite complex (a cascade of \texttt{exists} expressions, one for each clause). The \textit{direct} strategy looks at the source expression and performs the matching test directly on the clause patterns. If all possible values match one of the patterns, the obligation is probably provable. The other strategies would almost certainly return a \emph{MAYBE} status for the obligation.
\end{itemize}

\subsection{Polymorphic Functions}
\label{sec:polymorphic}

VDM can define polymorphic functions, and proof obligations generated from these functions may use the same polymorphic type parameters. For example:

\begin{verbatim}
functions
    f[@T]: seq of @T * nat -> @T
    f(s, i) == s(i);

> pog
Generated 1 proof obligation:

Proof Obligation 1: (Unproved)
f: sequence apply obligation in test.vdmsl at line 3:16
(forall s:seq of (@T), i:nat &
  i in set inds s)

> qc
PO #1, FAILED in 0.008s: Counterexample: i = 0, s = [],
  T = real
----
f: sequence apply obligation in test.vdmsl at line 3:16
(forall s:seq of (@T), i:nat &
  i in set inds s)
\end{verbatim}

\noindent
Notice that the proof obligation includes a \texttt{@T} type and the "qc" execution found a counterexample with the type parameter bound to \texttt{real}. However, this binding was not produced by the strategies, which are only capable of generating type binds (\ie variable name/value pairs).

By default, \emph{QuickCheck} will bind all type parameters to \texttt{real}. But in cases where this is not sensible, a \emph{@QuickCheck} annotation is available, which allows a list of sensible types to be specified for each polymorphic function. For example:

\begin{verbatim}
functions
    -- @QuickCheck @T = set of nat, set of bool;
    f[@T]: seq of @T * nat -> @T
    f(s, i) == s(i);

> qc
PO #1, FAILED in 0.002s: Counterexample: i = 0, s = [],
  T = set of (nat)
----
f: sequence apply obligation in test.vdmsl at line 4:16
(forall s:seq of (@T), i:nat &
  i in set inds s)
\end{verbatim}

\noindent
Now we see that the \texttt{@T} parameter is bound to \texttt{set of nat}, and both types in the \texttt{@QuickCheck} list would have been tried.

Quantifying over specification types sensibly is a complex area, and difficult to automate in the strategies, but this is potentially an area for future work.

\subsection{QuickCheck Performance}
\label{sec:performance}

\emph{QuickCheck} has been tested on a large number of legacy specifications, of all VDM dialects, that are distributed with the Overture tool. See Table~\ref{tab:results}.

\begin{table}
\centering

\begin{tabular}{|l |r |r |r |r |r|} \hline 
 & \textbf{VDM-SL} & \textbf{VDM++} & \textbf{VDM-RT} & \textbf{Totals} & \textbf{\%age} \\ \hline 
\#Specs& 50 & 51 & 13 & 114 &  \\ \hline 
\#POs& 4964 & 2830 & 435 & 8229 & \\ \hline 
PROVABLE  & 878 & 323 & 37 & 1238 & 15.04\% \\ \hline 
\textit{  - by trivial }& \textit{141} & \textit{91} & \textit{3} & \textit{235} & \textit{2.86\%} \\ \hline 
\textit{  - by finite }& \textit{227} & \textit{135} & \textit{16} & \textit{378} & \textit{4.59\%} \\ \hline 
\textit{  - by witness }& \textit{109} & \textit{30} & \textit{7} & \textit{146} & \textit{1.77\%} \\ \hline 
\textit{  - by direct }& \textit{401} & \textit{67} & \textit{11} & \textit{479} & \textit{5.82\%} \\ \hline 
MAYBE  & 2077 & 781 & 108 & 2966 & 36.04\% \\ \hline 
FAILED (counterexample) & 942 & 128 & 5 & 1075 & 13.06\% \\ \hline 
UNCHECKED  & 1057 & 1598 & 285 & 2940 & 35.73\% \\ \hline 
TIMEOUT (5s) & 10 & 0 & 0 & 10 & 0.12\% \\ \hline 
 \multicolumn{5}{|c|}{}& \textbf{100.00\%} \\ \hline

\end{tabular}
\smallskip
\caption{\emph{QuickCheck} results for Overture example specifications}
\label{tab:results}
\end{table}

We consider that \emph{PROVABLE} and \emph{FAILED} results are useful, and hence across all dialects, \emph{QuickCheck} has produced useful results for about 28\% of proof obligations. If we disregard the \emph{UNCHECKED} obligations, that figure rises to 41\%.

In terms of the evaluation performance on individual obligations, the time taken\footnote{On a 2.4GHz Core i7, with 32Gb RAM.} depends on the status returned\footnote{The variation in performance between dialects reflects the nature of the example specifications used, rather than being due to \emph{QuickCheck}.}. See Table~\ref{tab:performance}.

\begin{table}
\centering

\begin{tabular}{|l |r |r |r |r|} \hline 
 & \textbf{VDM-SL} & \textbf{VDM++} & \textbf{VDM-RT} & \textbf{Average (ms)} \\ \hline 
PROVABLE  & 4.1 & 4.4 & 1.7 & 3.3 \\ \hline 
FAILED  & 10.4 & 3.1 & 11.2 & 8.2 \\ \hline 
MAYBE  & 45.5 & 49.3 & 13.3 & 36.1 \\ \hline

\end{tabular}
\smallskip
\caption{\emph{QuickCheck} performance per obligation, in ms.}
\label{tab:performance}
\end{table}

Note that \emph{PROVABLE} obligations are generally the fastest. This is because when a strategy can claim provability, it does not need to generate any further potential counterexamples. \emph{FAILED} obligations are a bit slower, because these need to evaluate a potentially large number of values before the counterexample is found. Lastly, the \emph{MAYBE} case is typically the slowest, because it has to exhaust all of the possible values before concluding that it cannot be sure.

We consider that "a few milliseconds" per obligation is fast enough to justify the "Quick" title of the tool. In practice, this means that most specifications can have all their POs processed without a significant disruption to workflow.

\section{QuickCheck Tool Integration}
\label{sec:integration}

\emph{QuickCheck} is implemented as a \textit{plugin} in both the VDMJ command line and VDM-VSCode graphical environments. This means that a Java jar containing the tool simply has to be added to the classpath, and the features of the tool automatically become available (\ie new "qc" and "qr" commands appear, or new UI options are enabled). The core processing of the two plugins is identical, but there are some differences in a thin "wrapper" layer around that, to make the functionality available to the two environments.

\subsection{VDMJ Integration}
\label{sec:vdmj}

\emph{QuickCheck} is executed via the "quickcheck" command, which is usually abbreviated to "qc". The \texttt{-?} option to the command gives a listing of its options:

\begin{verbatim}
Usage: quickcheck [-?|-help][-q|-v][-t <secs>]
  [-i <status>]* [-s <strategy>]* [-<strategy:option>]*
  [<PO numbers/ranges/patterns>]

  -?|-help           - show command help
  -q|-v              - run with minimal or verbose output
  -t <secs>          - timeout in secs
  -i <status>        - only show this result status
  -s <strategy>      - enable this strategy (below)
  -<strategy:option> - pass option to strategy
  PO# numbers        - only process these POs
  PO# - PO#          - process a range of POs
  <pattern>          - process PO names or modules matching

Enabled strategies:
  fixed [-fixed:file <file> |
    -fixed:create <file>][-fixed:size <size>]
  search (no options)
  finite [-finite:size <size>]
  trivial (no options)
  direct (no options)

Disabled strategies (add with -s <name>):
  random [-random:size <size>][-random:seed <seed>]
>
\end{verbatim}

\noindent
Hopefully these options are intuitive. The output also lists the strategies that are enabled and disabled. By default, everything is enabled except the \textit{random} strategy\footnote{This is because the \texttt{random} strategy and the \texttt{fixed} strategy are very similar, but \texttt{fixed} usually produces better results.}, but this can be changed by using one or more \texttt{-s} options. And by default "qc" will process all of the obligations in the current module or class (changed with the \texttt{default} command).

Note that the strategies that generate a large number of possible values come with a \texttt{-<name>:size} option to override the default. Increasing this makes evaluation slower, but more likely to catch counterexamples.

The output of the "qc" command is a list of the checked status of each obligation processed, together with any counterexample or witness values found. For example:

\begin{verbatim}
> qc
PO #1, PROVABLE by direct (body is total) in 0.003s
PO #2, PROVABLE by witness c = 0, r = 0 in 0.0s
PO #3, PROVABLE by direct (body is total) in 0.001s
PO #4, PROVABLE by witness b = {} in 0.001s
PO #5, PROVABLE by finite types in 6.232s
PO #6, PROVABLE by finite types in 0.0s
PO #7, MAYBE in 0.001s
PO #8, MAYBE in 0.001s
PO #9, MAYBE in 0.104s
PO #10, UNCHECKED
PO #11, MAYBE in 0.007s
PO #12, TIMEOUT in 6.356s
PO #13, UNCHECKED
PO #14, UNCHECKED
PO #15, PROVABLE by direct (body is total) in 0.001s
>
\end{verbatim}

\noindent
A \emph{PROVABLE by witness} result is when a strategy has found a binding that makes an existential obligation true\footnote{And it ought to state the strategy used.}.

The "qcrun" command, usually abbreviated to "qr", is much simpler to use and can only be executed with one proof obligation number. Note that "qc" must have been executed first, in order to generate the counterexample needed for "qr" to attempt to debug the obligation.

\begin{verbatim}
> qr 1
Obligation does not have a counterexample/witness. Run qc?

> qc
PO #1, FAILED in 0.002s: Counterexample: i = 0, s = []
----
f: sequence apply obligation in test.vdmsl at line 3:16
(forall s:seq of nat, i:nat &
  i in set inds s)

> qr 1
=> print f([], 0)
Error 4064: Value 0 is not a nat1 in test.vdmsl at line 3:16
3:      f(s, i) == s(i);
MainThread>
\end{verbatim}

\noindent
Note that the "qr 1" command is equivalent to the "print f([], 0)" command shown, because those are the counterexample values found by "qc". Generally, "qr" tries to match the counterexample bindings to the enclosing function's parameters. With complex obligations this may not be possible.

\subsection{VDM-VSCode Integration}
\label{sec:vscode}
The VDM-VSCode integration provides a graphical alternative to the command line-based VDMJ interface. The integration is included with version 1.4.0 and newer of the VDM VSCode extension for Visual Studio Code. Using the integration requires no additional setup, as the \emph{QuickCheck} plugin comes bundled with the extension and the intricacies of adding the Java jar to the classpath are handled automatically.

\begin{figure}
    \centering
    \includegraphics[width=\linewidth]{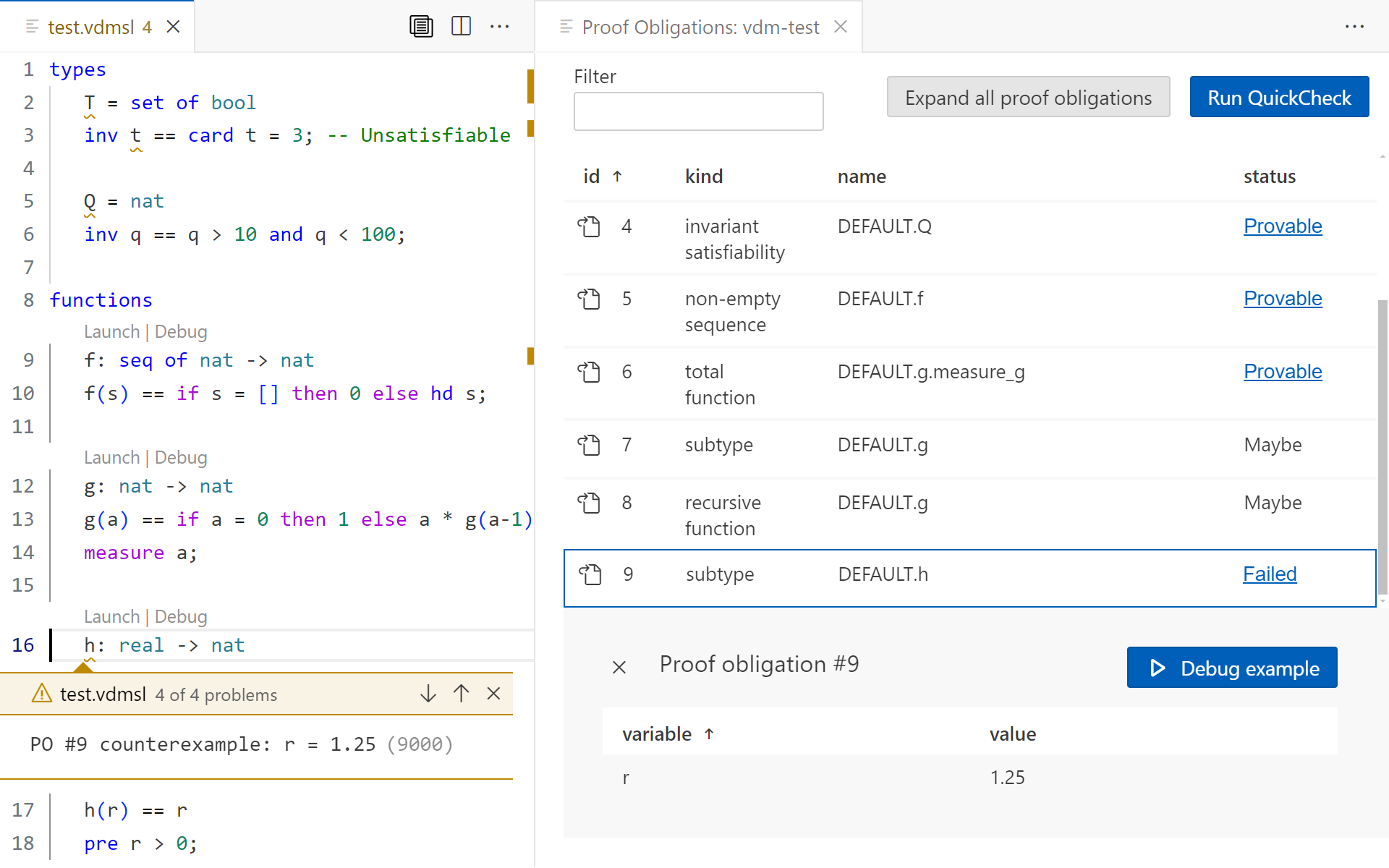}
    \caption{\textit{QuickCheck} in the VDM VSCode Extension.}
    \label{fig:qc_vs_code_ui}
\end{figure}

The Proof Obligation Generation (POG) view shown in Figure \ref{fig:qc_vs_code_ui} acts as the main point of interaction with \emph{QuickCheck}. It is from within this view that the tool is executed, where the results of the execution are presented, and where further debugging is initiated. As a rule of thumb, it usually only takes a few seconds to check all the obligations of a specification, with the exception of very complex specifications or when using enhanced strategies. During an active execution, the incremental progress of checking is displayed in a notification box, where the user is also able to cancel checking prematurely. To reduce the time required to check obligations, a search filter can be applied prior to executing \textit{QuickCheck}, since only the visible subset of proof obligations are checked. Upon successful execution of \emph{QuickCheck}, the statuses of the proof obligations are updated in the view, possibly turning into links. Clicking on one of these links, opens up the \textit{QuickCheck Panel} at the bottom of the PO View, showing supplemental information about the associated proof obligation, including the strategy used and an optional error message. If \emph{QuickCheck} found counterexample or witness values, these are shown in a table in the \textit{QuickCheck Panel} along with a button to start a debug session with these values bound to the identifiers listed in the table.

\emph{QuickCheck} is optionally configurable via a file named ``quickcheck.json'' in the ``.vscode'' directory at the root of the project. If no such file exists, \emph{QuickCheck} falls back to a set of sensible defaults. If the configuration file does exist, but leaves out certain settings, these take on their default values. The following example configuration sets the timeout to 10 seconds, overriding the default of 1 second, and configures the strategies to use:

\begin{verbatim}
{
  "config": {
    "timeout": 10
  },
  "strategies": [
    {
      "name": "fixed",
      "enabled": true,
      "size": 1000
    },
    {
      "name": "direct",
      "enabled": false
    },
    {
      "name": "trivial",
      "enabled": true
    },
    {
      "name": "search",
      "enabled": false
    }
  ]
}
\end{verbatim}

\noindent
It is worth noting that some strategies take additional parameters aside from the common ``name'' and ``enabled'' parameters. Looking at the configuration of the ``fixed'' strategy above, the ``size'' parameter stands out as one of these strategy-specific configurations. It is up to the implementer of the strategy to decide on the interpretation of the configuration.

\section{Related Work}
\label{sec:related}

\emph{QuickCheck} is inspired by the testing tool for Haskell \cite{Claessen&00,Bird98}. This approach has been adopted by other verified language environments, such as Isabelle \cite{Nipkow89b,Nipkow&02}, that use a library called ``QuickCheck''~\cite{Bulwahn12}.

\section{Future Work}
\label{sec:future}

The \emph{QuickCheck} tool is effectively a "hook" into VDMJ which allows further proof features to be added, initially as extra pluggable strategies (see \ref{sec:design}). The following ideas may be worth exploration:

\begin{itemize}
\item \emph{Integration with an SMT solver.} If proof obligations can be translated into Dafny, or SMTLIB directly, it would be possible for SMT solvers like Z3 or CVC to work on the discharge of obligations via a \emph{QuickCheck} plugin.
\item\emph{Improved analysis for UNCHECKED operation POs.} We need to improve the quality of POs generated from operations to allow them to be checked. This will be a challenge for the object oriented dialects.
\item \emph{Using AI to find counterexamples.} Proof obligations have a very regular form, because they are predictable consequences of the original VDM source. It may therefore be possible to train a machine learning model to recognise obligation patterns and suggest counterexamples, if we can find enough training data.
\item \emph{Quantification over polymorphic types.} As mentioned in Section \ref{sec:polymorphic}, it is difficult for \emph{QuickCheck} to automatically quantify over polymorphic types, especially given the context of a particular specification. This would require a deeper analysis of the specification to choose sensible type parameters.
\end{itemize}

\newpage
\paragraph{\textbf{Acknowledgements}}
We are grateful for the support of the European Union, Newcastle University, Aarhus University and the Poul Due Jensen foundation. In particular, we thank Leo Freitas and Peter Gorm Larsen for their invaluable help on testing early versions of the tool. We also thank the reviewers for valuable feedback on the original version of this paper.

\bibliographystyle{splncs04}

\end{document}